\documentclass[aip,amsmath,reprint]{revtex4-1} 

\usepackage{bm}
\usepackage{graphicx}
\usepackage{mathptmx}
	
\begin{document}

\title{Kinetic inductance of superconducting nanostrips with turns}

\author{Yasunori Mawatari}
\affiliation{%
National Institute of Advanced Industrial Science and Technology (AIST), \\
Tsukuba, Ibaraki 305--8568, Japan
}

\date{12 May 2023}

\begin{abstract}
Kinetic inductances of superconducting nanostrips with a meander pattern are theoretically investigated based on the London model, and the effect of the current crowding at the turns of the nanostrips is considered. 
The complex current approach is developed for analytical investigation of the kinetic inductance of nanostrips with turns for thin $d<\lambda$ and narrow $w\ll \lambda^2/d$ superconducting strips, where $d$ is the strip thickness, $w$ is the strip width, and $\lambda$ is the London penetration depth. 
We show that the current distribution in superconducting nanostrips of $wd\ll\lambda^2$ is identical to that in normal conducting nanostrips of $wd\ll\delta^2/2$, where $\delta$ is the skin depth, and the dependence of the kinetic inductance on the nanostrip geometry is identical to that of the normal resistance. 
Effects of the edge defects of superconducting strips upon the kinetic inductance are also considered. 
\end{abstract}


\maketitle 

\section{Introduction} 
Superconducting nanostrip single-photon detectors (SSPDs) have been extensively developed for applications in optical communications, quantum information, and optical quantum computing.~\cite{Natarajan_12} 
The reset time of SSPDs is proportional to the kinetic inductance of superconducting nanostrips,~\cite{Kerman_06} which is a critical parameter for realizing SSPDs with a fast response. 
The kinetic inductance is crucial in microwave kinetic inductance detectors.~\cite{Day_03} 
In single flux quantum (SFQ) circuits, the implementation of the kinetic inductors can considerably reduce the area occupied by the cell inductors.~\cite{Tolpygo_16} 

Typical SSPDs are fabricated by patterning superconducting films with meander nanostrips with many turns, and Clem \textit{et al}.~\cite{Clem_11,Clem_12a} theoretically investigated the effects of current crowding on the critical currents of superconducting nanostrips with turns. 
However, the current crowding effects on the kinetic inductance have not yet been clarified.
Although it is possible to numerically estimate the kinetic inductance, simple analytical formulas are useful in the early stage of designing superconducting electronic devices. 

In this paper, we theoretically investigate the two-dimensional current distribution and kinetic inductance of superconducting nanostrips with meandering turns. 
This paper is organized as follows: 
basic equations based on the London model for analyzing the current distribution and kinetic inductance are shown in Sec.~\ref{sec:basic-eqs}; 
analytical expressions for the current distribution and the kinetic inductance of meander superconducting nanostrips are derived in Sec.~\ref{sec:meander}; 
and our results are summarized in Sec.~\ref{sec:summary}.

\section{Basic equations for superconducting nanostrips
\label{sec:basic-eqs}} 
\subsection{Kinetic inductance based on the London model} 
The London model is valid for superconductors of $\lambda/\xi\gg 1$, where $\lambda$ is the London penetration depth and $\xi$ is the coherence length. 
The London free energy $F_\textrm{London}= F_\textrm{m} +F_\textrm{k}$ is composed of the magnetic energy, $F_\textrm{m}= \int(|\bm{B}|^2/2\mu_0) dV$, and the kinetic energy, $F_\textrm{k}= \int(\mu_0\lambda^2|\bm{J}|^2/2) dV$, where $\bm{B}$ is the magnetic field, $\bm{J}$ is the current density, and $\mu_0$ is the vacuum permeability. 
For a superconductor carrying transport current $I_0$, the magnetic and kinetic energy can be rewritten as $F_\textrm{m}= L_\textrm{m}I_0^2/2$ and $F_\textrm{k}= L_\textrm{k}I_0^2/2$, respectively, where $L_\textrm{m}$ is the magnetic inductance and $L_\textrm{k}$ is the kinetic inductance.~\cite{Meservey_69,Clem_13,Yoshida_92,Annunziata_10} 

In this paper, we investigate thin superconducting nanostrips of thickness $d$ and width $w$ in the $xy$-plane, where $d<\lambda$ and $w\ll \lambda^2/d$. 
In these thin narrow strips, kinetic energy is dominant, $F_\textrm{London}\simeq F_\textrm{k}\gg F_\textrm{m}$, and the kinetic inductance is dominant, $L_0\simeq L_\textrm{k}\gg L_\textrm{m}$. 
Thus, we consider the kinetic inductance (in units of [H]) of superconducting nanostrips defined by 
\begin{equation}
	L_\textrm{k} =\frac{\mu_0\lambda^2}{I_0^2}\int |\bm{J}|^2 dV
	= \frac{\mu_0\lambda^2 d}{I_0^2}\iint \left( J_x^2+J_y^2 \right) dxdy, 
\label{eq:Lk}
\end{equation}
where $J_x$ and $J_y$ are the components of the current density in the $xy$-plane. 
For a straight nanostrip with no turns of thickness $d$ and width $w$, the current density is uniform, $J_x= I_0/wd$ and $J_y=0$, and the kinetic inductance, $L_\textrm{k}=L_\textrm{k0}$, calculated from Eq.~\eqref{eq:Lk} is given by 
\begin{equation}
	L_\textrm{k0} =\frac{\mu_0\lambda^2}{d} \frac{\ell_0}{w}, 
\label{eq:Lk0}
\end{equation}
where $\ell_0$ is the length of a straight nanostrip.

\subsection{Normal resistance and kinetic inductance} 
The dissipation power of superconducting nanostrips in the normal state is given by $R_\textrm{n} I_0^2 =\int\rho_\textrm{n} |\bm{J}|^2 dV$, where $R_\textrm{n}$ is the normal resistance (in units of [$\Omega$]) and $\rho_\textrm{n}$ is the normal resistivity of nanostrips. 
Thus, the normal resistance is given by 
\begin{equation}
	R_\textrm{n} =\frac{\rho_\textrm{n}}{I_0^2} \int |\bm{J}|^2 dV 
	= \frac{\rho_\textrm{n} d}{I_0^2}\iint \left( J_x^2+J_y^2 \right) dxdy. 
\label{eq:Rn}
\end{equation}
Although the expression $L_\textrm{k}/\mu_0\lambda^2=\int|\bm{J}/I_0|^2 dV$ of Eq.~\eqref{eq:Lk} is similar to $R_\textrm{n}/\rho_\textrm{n}=\int|\bm{J}/I_0|^2 dV$ of Eq.~\eqref{eq:Rn}, the volume integral of $\int|\bm{J}/I_0|^2 dV$ in superconductors does not necessarily coincide with that in normal conductors. 
As shown in Appendix~\ref{app:current-nanostrips}, however, the current distribution in superconducting nanostrips of $wd\ll\lambda^2$ is identical to that in normal conducting nanostrips of $wd\ll\delta^2/2$, where $\delta$ is the skin depth. 
The volume integral $\int|\bm{J}/I_0|^2 dV$ for such nanostrips is determined by the geometry of nanostrips and is independent of the material parameters. 

Equations~\eqref{eq:Lk} and \eqref{eq:Rn} are, therefore, rewritten as 
\begin{equation}
	\frac{L_\textrm{k}}{\mu_0\lambda^2} =\frac{R_\textrm{n}}{\rho_\textrm{n}},
\label{eq:Lk/Rn}
\end{equation}
which means that the kinetic inductance is proportional to the normal resistance, regardless of the geometrical structure of the superconducting nanostrips. 
Equation~\eqref{eq:Lk/Rn} has already been suggested in Refs.~\onlinecite{Yoshida_92,Kerman_06} without careful consideration of the identity of the current distribution in superconductors and that in normal conductors.

Hall~\cite{Hall_68} theoretically investigated the normal resistance of thin-film patterns (i.e., strips with turns) of various geometries. 
Most of the results of the normal resistance shown in Ref.~\onlinecite{Hall_68} are expressed as the general form, $R_\textrm{n}=(\rho_\textrm{n}/d)[(\ell_0/w) +\alpha]$, where $\alpha$ is a numerical factor arising from the effect of turns. 

Because of the proportional relationship, $L_\textrm{k}\propto R_\textrm{n}$, in Eq.~\eqref{eq:Lk/Rn}, analytical formulas for $L_\textrm{k}$ of superconducting nanostrips are simply obtained by converting the formulas for $R_\textrm{n}$ listed in Ref.~\onlinecite{Hall_68}, as suggested by Tolpygo {\it et al}.~\cite{Tolpygo_22,Tolpygo_23} 
Some of the $L_\textrm{k}$ formulas for typical geometries are listed in the Appendix \ref{app:Lk-turns}. 
The general expression of the kinetic inductance for $\ell_0\gg w$ is also given by 
\begin{equation}
	L_\textrm{k}= \frac{\mu_0\lambda^2}{d} \left(\frac{\ell_0}{w} +\alpha\right),  
\label{eq:Lk_general}
\end{equation}
which is similar to Eq.~\eqref{eq:Lk0} for straight strips, although correction factor $\alpha$ arises from the effect of turns. 
Numerical factor $\alpha$ in Eq.~\eqref{eq:Lk_general} depends on the geometry of the turns of nanostrips and tends to be large when the current crowding effect is large.

\subsection{Complex current density and kinetic inductance} 
We investigate the two-dimensional current density, ${\bm J}= J_x(x,y)\hat{\bm x} +J_y(x,y)\hat{\bm y}$, in the $xy$-plane based on the London model, by using the complex current approach.~\cite{Hagedon_63,Hall_68,Clem_11,Clem_12a} 

The current density is solenoidal, $\nabla\cdot{\bm J}=0$, because of Amp\`{e}re's law, ${\bm J}= \nabla\times{\bm B}/\mu_0$, and we have ${\bm J}=\nabla\times(\psi\hat{\bm z})$, where $\psi(x,y)$ is a scalar function. 
The London equation in superconductors without vortices, $\nabla\times{\bm J}= -{\bm B}/\mu_0\lambda^2$, can be approximated as $\nabla\times{\bm J}=0$ for nanostrips of $w\ll \lambda^2/d$ with no applied magnetic field, as shown in Appendix~\ref{app:current-nanostrips}, and we have ${\bm J}= -\nabla\varphi$, where $\varphi(x,y)$ is a scalar function. 
This approximation is consistent with the neglect of the magnetic energy, as in Sec.~\ref{sec:basic-eqs}.~A. 
See Ref.~\onlinecite{Via_13} for a precise numerical investigation considering the magnetic energy. 

The two-dimensional current density is, therefore, expressed as 
\begin{align}
	J_x &= \partial\psi/\partial y =-\partial\varphi/\partial x, 
\label{eq:Jx}\\
	J_y &= -\partial\psi/\partial x =-\partial\varphi/\partial y. 
\label{eq:Jy}
\end{align}
Equations~\eqref{eq:Jx} and \eqref{eq:Jy} correspond to the Cauchy--Riemann conditions~\cite{Arfken_01} for $\psi$ and $\varphi$. 
Therefore, complex stream function $\mathcal{G}$ and complex current $\mathcal{J}$ defined by 
\begin{align}
	\mathcal{G} &= \psi +i \varphi,
\label{eq:cG}\\
	\mathcal{J} &=J_y+iJ_x =-d\mathcal{G}/d\zeta 
\label{eq:cJ}
\end{align}
are analytic functions of complex variable $\zeta= x+iy$. 
The streamlines of the current density correspond to the contour lines of $\psi=\textrm{Re}\,\mathcal{G}$. 
As shown in Appendix \ref{app:I0-Lk-phi-phi}, transport current $I_0$ and kinetic inductance $L_\textrm{k}$ defined by Eq.~\eqref{eq:Lk} are related to $\psi$ and $\varphi$ by 
\begin{align}
	I_0/d &= \psi_2-\psi_1, 
\label{eq:I0-psi}\\
	L_\textrm{k}I_0/\mu_0\lambda^2 &= \varphi_2-\varphi_1, 
\label{eq:Lk_phi}
\end{align}
where $\psi_1$ and $\psi_2$ are the values of $\psi$ at the sides and $\varphi_1$ and $\varphi_2$ are the values of $\varphi$ at the ends of the nanostrips.

\section{Meander superconducting nanostrips 
\label{sec:meander}} 
In this section, we investigate the current distribution and kinetic inductance of meander superconducting nanostrips. 
The formulas presented in this section are similar to those for magnetic field distribution and effective permeability of a hexagonal array of superconducting strips.~\cite{Mawatari_12,Mawatari_13} 

\subsection{Current distribution in meander nanostrips} 
\begin{figure}[b]
\includegraphics{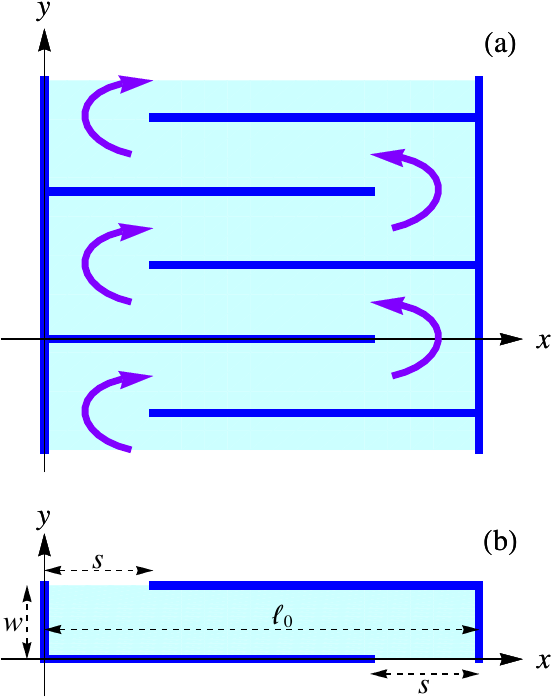}
\caption{%
Geometry of a meander superconducting nanostrip in the $xy$-plane.
The thickness of the edges of the nanostrips (thick lines) is infinitesimal. 
(a) Periodic meander structure extended in the $y$-direction. The thick arrows show the schematic direction of the current flow. 
(b) Unit cell of $0<x<\ell_0$ and $0<y<w$, where $\ell_0$ is the strip length, $w$ is the strip width, and $s$ is the turn width. 
The variable transform of Eq.~\eqref{eq:eta-zeta} maps the points $\zeta=0,\ \ell_0 -s,\ \ell_0,\ \ell_0+iw,\ s+iw$, and $iw$ in the $\zeta$-plane onto the points $\eta=0,\ \gamma,\ 1,\ 1/k_1,\ \beta$, and $\infty$ in the $\eta$-plane, respectively. 
}
\label{fig:meander}
\end{figure}

We investigate a model of meander superconducting nanostrips, the geometry of which is shown in Fig.~\ref{fig:meander} (a). 
Because of the periodic structure along the $y$-direction, it is sufficient to consider a unit cell of length $\ell_0$ and width $w$ (i.e., $0<x<\ell_0$ and $0<y<w$) [Fig.~\ref{fig:meander} (b)]. 

Here we introduce the conformal mapping from the $\zeta=x+iy$-plane to the $\eta=u+iv$-plane,~\cite{Mawatari_12,Mawatari_13} 
\begin{equation}
	\eta(\zeta)= \mbox{sn}(\zeta/c,k_1), 
\label{eq:eta-zeta}
\end{equation}
where $\mbox{sn}(u,k)$ is the sine amplitude (i.e., the Jacobi sn function).~\cite{Gradshtein_94} 
Modulus $k_1$ in Eq.~\eqref{eq:eta-zeta} is determined as the function of $\ell_0/w$ by solving 
\begin{equation}
	\frac{\ell_0}{w}= \frac{\textbf{K}(k_1)}{\textbf{K}\left(\sqrt{1-k_1^2}\right)},  
\label{eq:k1}
\end{equation}
where $\textbf{K}(k)$ is the complete elliptic integral of the first kind.~\cite{Gradshtein_94} 
Parameter $c$ in Eq.~\eqref{eq:eta-zeta} is given by 
\begin{equation}
	c = \frac{\ell_0}{\textbf{K}(k_1)} = \frac{w}{\textbf{K}\left(\sqrt{1-k_1^2}\right)}. 
\label{eq:c}
\end{equation}
The unit cell in the $\zeta=x+iy$-plane, $0<x<\ell_0$ and $0<y<w$, is mapped onto the first quadrant of the $\eta=u+iv$-plane, $u>0$ and $v>0$. 

Using the conformal mapping technique, we obtain the complex current defined by Eq.~\eqref{eq:cJ}  as~\cite{Mawatari_12,Mawatari_13} 
\begin{equation}
	\mathcal{J}(\zeta)= J_0 \frac{\eta\sqrt{\eta^2-k_1^{-2}}}{%
		\sqrt{(\eta^2-\gamma^2) (\eta^2-\beta^2)}}, 
\label{eq:compex-J}
\end{equation}
where $\gamma$ and $\beta$ are defined by 
\begin{align}
	\gamma &= \eta(\ell_0-s) =\mbox{sn}((\ell_0-s)/c,k_1), 
\label{eq:gamma}\\
	\beta &= \eta(s+iw) =\sqrt{\frac{k_1^{-2}-\gamma^2}{1-\gamma^2}}. 
\label{eq:beta}
\end{align}
Parameter $J_0=J_y(0,w)$ in Eq.~\eqref{eq:compex-J} is the current density at $(x,y)=(0,w)$, and $J_0$ is proportional to transport current $I_0$, as shown below. 

The complex stream function, $\mathcal{G}=\int_{iw}^{\zeta}\mathcal{J}(\zeta')d\zeta'$, is obtained by integrating Eq.~\eqref{eq:compex-J} as 
\begin{align}
	\mathcal{G}(\zeta) 
	&= \frac{cJ_0}{k_1} \int_{\eta}^{\infty} 
		\frac{\eta'd\eta'}{\sqrt{(\eta'^2-1)(\eta'^2-\beta^2)(\eta'^2-\gamma^2)}} 
\nonumber\\
	&= \frac{cJ_0}{k_1\sqrt{\beta^2-\gamma^2}}\ 
		F\left(\arcsin \sqrt{\frac{\beta^2-\gamma^2}{\eta^2-\gamma^2}},\ k_2\right), 
\label{eq:complex-G}
\end{align}
where $F(\theta,k)$ is the elliptic integral of the first kind,~\cite{Gradshtein_94} and the modulus, $k_2$, is defined by
\begin{equation}
	k_2= \left[ 1+\frac{k_1^{-2}-1}{(1-\gamma^2)^2} \right]^{-1/2}. 
\label{eq:k2}
\end{equation}
Coefficient $J_0$ in Eqs.~\eqref{eq:compex-J} and \eqref{eq:complex-G} is determined by Eq.~\eqref{eq:I0-psi} and \eqref{eq:complex-G} as 
\begin{align}
	I_0/d 
	&= \psi(\ell_0,0)-\psi(\ell_0-s,0) 
\nonumber\\
	&= \textrm{Re}\,[\mathcal{G}(\ell_0) -\mathcal{G}(\ell_0-s)] 
\nonumber\\
	&= \frac{cJ_0}{k_1} \int_{\gamma}^1 
		\frac{udu}{\sqrt{(u^2-\gamma^2)(1-u^2)(\beta^2-u^2)}} 
\nonumber\\
	&= \frac{cJ_0}{k_1\sqrt{\beta^2-\gamma^2}} \textbf{K}(k_2).
\label{eq:I0-J0}
\end{align}

\begin{figure*}[bth]
 \includegraphics{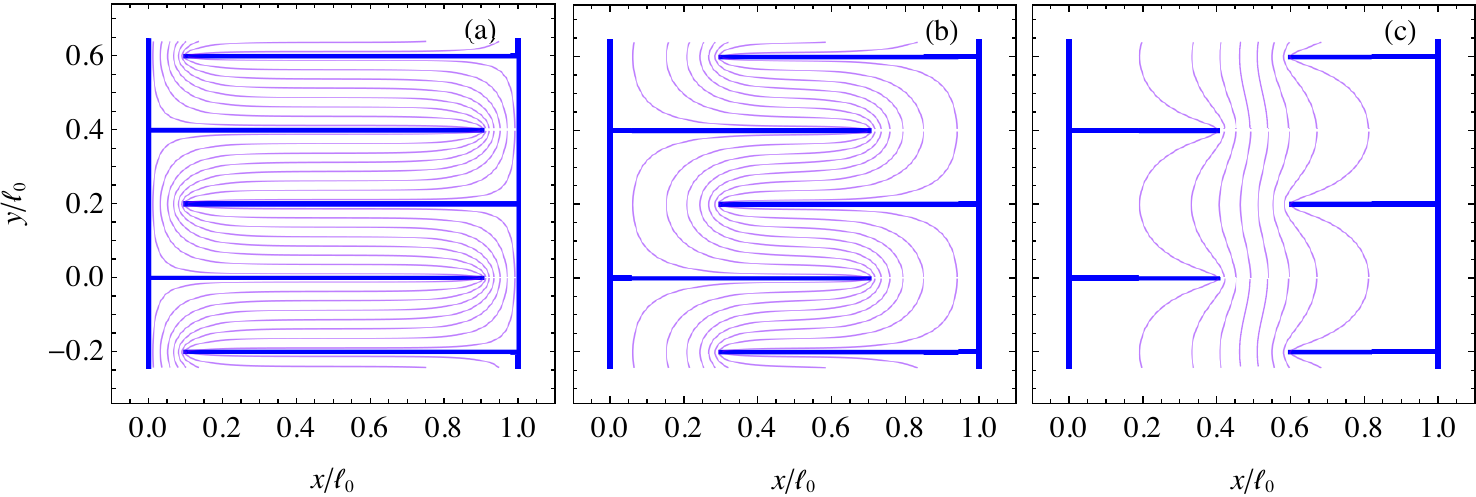}
\caption{%
Streamlines of the current density in meander superconducting nanostrips of $w/\ell_0=0.2$ for (a) $s/\ell_0=0.1\ (s< w)$, (b) $s/\ell_0=0.3\ (w<s<\ell_0/2)$, and (c) $s/\ell_0=0.6\ (s>\ell_0/2)$. 
Thick lines correspond to the edges of the nanostrips.  
}
\label{fig:current-flow}
\end{figure*}

Figure~\ref{fig:current-flow} shows the streamlines of the current density, which correspond to the contour lines of $\psi(x,y)=\textrm{Re}\,\mathcal{G}(x+iy)$ obtained from Eq.~\eqref{eq:complex-G}.

\subsection{Kinetic inductance of meander nanostrips} 
\begin{figure}[b]
\includegraphics{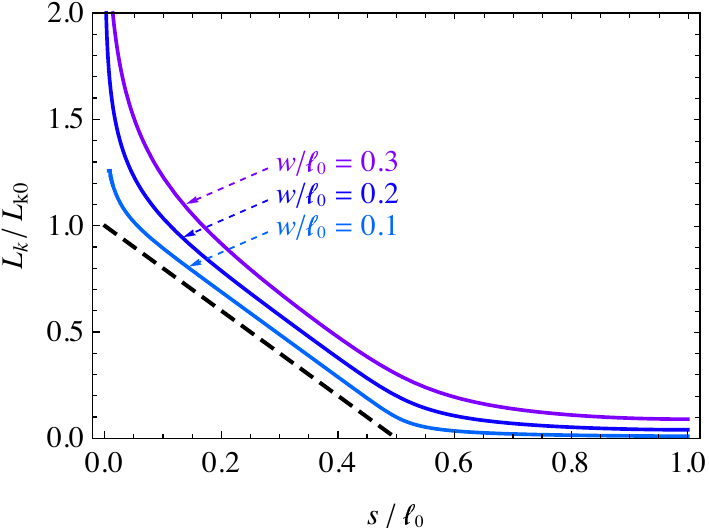}
\caption{%
Kinetic inductance $L_\textrm{k}$ vs turn width $s$ for $w/\ell_0=0.1$, $0.2$, and $0.3$. 
The dashed line corresponds to the limit of $w/\ell_0\to 0$, $L_\textrm{k}=\mu_0\lambda^2 (\ell_0-2s)/(wd)$. 
$L_\textrm{k}$ and $s$ are normalized by $L_\textrm{k0}$ from Eq.~\eqref{eq:Lk0} and $\ell_0$, respectively.
}
\label{fig:Lk-s}
\end{figure}

Here, we consider the kinetic inductance, $L_\textrm{k}$, for one unit cell of the superconducting nanostrips [Fig.~\ref{fig:meander}(b)]. 
Using Eqs.~\eqref{eq:Lk_phi} and \eqref{eq:complex-G}, we have
\begin{align}
	L_\textrm{k}I_0/\mu_0\lambda^2 
	&= \varphi(\ell_0,0)-\varphi(s,w) 
\nonumber\\
	&= \textrm{Im}\,[\mathcal{G}(\ell_0) -\mathcal{G}(s+iw)] 
\nonumber\\
	&= \frac{cJ_0}{k_1} \int_1^{\beta} 
		\frac{udu}{\sqrt{(u^2-\gamma^2)(u^2-1)(\beta^2-u^2)}} 
\nonumber\\
	&=  \frac{cJ_0}{k_1\sqrt{\beta^2-\gamma^2}} 
		\textbf{K}\left(\sqrt{1-k_2^2}\right). 
\label{eq:Lk-calc}
\end{align}
Substituting Eq.~\eqref{eq:I0-J0} into Eq.~\eqref{eq:Lk-calc} yields the kinetic inductance, 
\begin{equation}
	L_\textrm{k} = \frac{\mu_0\lambda^2}{d} \frac{\textbf{K}\left(\sqrt{1-k_2^2}\right)}{\textbf{K}(k_2)} .  
\label{eq:Lk_result}
\end{equation}
Figure~\ref{fig:Lk-s} shows the plot of $L_\textrm{k}$ as functions of $s/\ell_0$ obtained from Eqs.~\eqref{eq:Lk_result}. 

In Eqs~\eqref{eq:Lk_s-narrow}--\eqref{eq:Lk_s-wide}, we demonstrate simple formulas for $L_\textrm{k}$ as functions of turn width $s$ for narrow strips of $w\ll \ell_0$. 
See Appendix \ref{app:Lk-simple} for the derivation of those equations.  
\begin{align}
	L_\textrm{k} &\simeq \frac{\mu_0\lambda^2}{d} 
		\left[\frac{\ell_0}{w}+\frac{4}{\pi}\ln\left(\frac{2w}{\pi s}\right)\right] 
\quad\text{for } s\ll w, 
\label{eq:Lk_s-narrow}\\
	L_\textrm{k} &\simeq \frac{\mu_0\lambda^2}{d} 
		\left[\frac{\ell_0-2s}{w}+\frac{4}{\pi}\ln(2)\right] 	
\quad\text{for } w<s<\ell_0/2, 
\label{eq:Lk_s-medium}\\
	L_\textrm{k} &\simeq \frac{\mu_0\lambda^2}{d} 
		\left[\frac{2s-\ell_0}{w}+\frac{4}{\pi}\ln(2)\right]^{-1} 
\ \text{for } \ell_0/2<s<\ell_0.  
\label{eq:Lk_s-wide}
\end{align}
For the case where turn width $s$ is much smaller than strip width $w$ (i.e., $s\ll w$), current crowding occurs at the turns [Fig.~\ref{fig:current-flow}(a)], resulting in large kinetic inductance [Eq.~\eqref{eq:Lk_s-narrow} and Fig.~\ref{fig:Lk-s}]. 
For the most realistic case for SSPDs, $w<s<\ell_0/2$, current is not so crowded at the turns [Fig.~\ref{fig:current-flow}(b)], and the effective strip length is given by $\ell_0-2 s$ 
[Eq.~\eqref{eq:Lk_s-medium} and the dashed line in Fig.~\ref{fig:Lk-s}].
We see a noticeable increase of $L_\textrm{k}$ for $w/\ell_0>0.1$ especially for $s<w$, and Fig.~\ref{fig:Lk-s} is useful to estimate $L_\textrm{k}$ for devices using superconducting strips with $w/\ell_0>0.1$. 

Although the case where $s>\ell_0/2$ is unrealistic for SSPDs, Eq.~\eqref{eq:Lk_s-wide} is useful to consider the effects of the periodic edge defects in straight strips [i.e., a strip of width $\ell_0$ with the notches of length $\ell_0-s$ and of periodicity $w$, as seen in Fig.~\ref{fig:current-flow}(c)]. 
The effective width of the current flow along the $y$ axis is about $2s-\ell_0$, which is smaller than the actual width $\ell_0$, as seen in Eq.~\eqref{eq:Lk_s-wide} for $\ell_0-s\gg w$. 
For small-defect limit of $\ell_0-s\ll w$, we have $L_\textrm{k}\simeq (\mu_0\lambda^2/d)[(w/\ell_0) +(\pi/2)(1-s/\ell_0)^2]$.

\section{Discussion and summary
\label{sec:summary}} 
On the basis of the London model, we theoretically investigate the current flow in superconducting nanostrips with meander turns, using the complex current approach and conformal mapping. 
The simple relationships between current, kinetic inductance, and the complex stream function given by Eqs.~\eqref{eq:I0-psi} and \eqref{eq:Lk_phi} are valid for any geometry of superconducting nanostrips and are useful for the analytical investigation of $L_\textrm{k}$ for more complicated geometry. 

In most long and narrow (i.e., $w\ll \ell_0$ and $w\ll \lambda^2/d$) superconducting nanostrips, the kinetic inductance is generally given by Eq.~\eqref{eq:Lk_general}, and factor $\alpha\sim 1$ can be neglected when the current crowding is nonsignificant. 
Although the critical currents of superconducting nanostrips are strongly affected by the current crowding at turns,~\cite{Clem_11,Clem_12a} its effects can be neglected in the kinetic inductance.~\cite{Tolpygo_23} 
For superconducting devices with nanostrips of $w/\ell_0>0.1$ the current crowding effect on $L_\textrm{k}$ is evident especially for $s<w$, as shown in Fig.~\ref{fig:Lk-s}. 

The current distribution in superconducting nanostrips of $wd\ll\lambda^2$ is identical to that in normal conducting nanostrips of $wd\ll\delta^2/2$, when the conductor geometry is identical. 
Because of the proportional relationship between the normal resistance $R_\textrm{n}$ of $wd\ll\delta^2/2$ and the kinetic inductance $L_\textrm{k}$ of $wd\ll\lambda^2$, the analytical formulas of $R_\textrm{n}$ for various geometries listed in Ref.~\onlinecite{Hall_68} can be used to estimate $L_\textrm{k}$. 

Equation~\eqref{eq:Lk_s-wide} for $\ell_0/2<s<\ell_0$ and $\ell_0-s\gg w$ demonstrates the kinetic inductance of a straight strip of width $\ell_0$ with edge defects of length $\ell_0-s$ and of periodicity $w$. 
The edge defects of small periodicity $w\ll \ell_0-s$ reduce the effective width $2s-\ell_0\,(<\ell_0)$ of the current flow, resulting in the increase in the kinetic inductance. 

The nonlinear microwave response of superconducting nanostrips and the quadratic current dependence of the kinetic inductance are also important problems in microwave applications of superconductors.~\cite{Lancaster_97,Day_03} 
We restrict the consideration to the case where the current is sufficiently smaller than the critical current and the liner London model is valid. 
The nonlinear response can be considered based on, for example, Ginzburg--Landau theory~\cite{Lam_92,Enpuku93,Cho_98} or microscopic theory.~\cite{Dam_96,Dam_97,Clem_12b} 
The theoretical investigation of the effects of turns in superconducting nanostrips on the nonlinear kinetic inductance will be addressed in future works.

\appendix
\section{Current density in nanostrips
\label{app:current-nanostrips}} 
We consider the response of superconducting nanostrips and of normal conducting nanostrips to the ac transport current with frequencies ranging from dc to at most microwave frequencies. 
We can safely disregard the displacement current in such a frequency regime, and Amp\`{e}re's law without the displacement current ${\bm J}= \nabla\times{\bm B}/\mu_0$ reduces to $\nabla\cdot{\bm J}=0$. 

In superconductors without an applied magnetic field, the current density follows the London equation given by $\nabla\times{\bm J}= -{\bm B}/\mu_0\lambda^2$, where the only magnetic field contribution is the self-field $\bm{B}=\bm{B}_1$ due to the transport current in strips. 
The solution of the London equation is given by ${\bm J}={\bm J}_0+{\bm J}_1$, where ${\bm J}_0$ is the solution neglecting the self-field, $\nabla\times{\bm J}_0= 0$, and ${\bm J}_1$ is the contribution from the self-field, $\nabla\times{\bm J}_1= -{\bm B}_1/\mu_0\lambda^2$. 
We estimate $|{\bm J}_1|$ comparing $|{\bm J}_0|$ by the perturbative calculation for $wd\ll \lambda^2$. 
The self-field due to ${\bm J}_0$ is roughly estimated as $|{\bm B}_1| \sim \mu_0|{\bm J}_0|d$,~\cite{Brandt_93,Zeldov_94} and we have $|{\bm J}_1| \sim |{\bm B}_1|w/\mu_0\lambda^2 \sim |{\bm J}_0|wd/\lambda^2$, because $|\nabla\times{\bm J}_1| \sim |{\bm J}_1|/w$. 
We see that the contribution from the self-field $|{\bm J}_1| \sim |{\bm J}_0|wd/\lambda^2 \ll |{\bm J}_0|$ can be neglected for $wd\ll \lambda^2$ in superconducting nanostrips. 

In normal conductors with the normal resistivity $\rho_\textrm{n}$ the current density $\bm{J}=\bm{E}/\rho_\textrm{n}$ follows the Faraday's law, $\nabla\times{\bm E}= -\partial {\bm B}/\partial t$ or $\nabla\times{\bm J}= -(1/\rho_\textrm{n})\partial {\bm B}_1/\partial t$. 
The solution in normal conductors is also given by ${\bm J}={\bm J}_0+{\bm J}_1$, where ${\bm J}_0$ is the solution neglecting the self-field, and ${\bm J}_1$ is the contribution from the self-field, $\nabla\times{\bm J}_1= -(1/\rho_\textrm{n})\partial {\bm B}_1/\partial t$. 
Because $|\partial {\bm B}_1/\partial t| \sim \mu_0\omega d|\bm{J}_0|$, we estimate $|{\bm J}_1| \sim |{\bm J}_0|\mu_0\omega wd/\rho_\textrm{n} =|{\bm J}_0|2wd/\delta^2$, where $\omega$ is the angular frequency and $\delta=\sqrt{2\rho_\textrm{n}/\mu_0\omega}$ is the skin depth. 
The contribution from the self-field $|{\bm J}_1| \sim |{\bm J}_0|2wd/\delta^2 \ll |{\bm J}_0|$ can be neglected for $wd\ll \delta^2/2$ in normal conducting nanostrips. 

The current density follows $\nabla\times{\bm J}=0$ neglecting the self-field effect in superconducting nanostrips of $wd\ll \lambda^2$ and also in normal conducting nanostrips of $wd\ll \delta^2/2=\rho_\textrm{n}/\mu_0\omega$. 
The current density following $\nabla\times{\bm J}=0$ and $\nabla\cdot{\bm J}=0$ is simply obtained as ${\bm J}=-\nabla\varphi$ by solving $\nabla^2\varphi=0$ with boundary conditions. 
We, thus, come to the conclusion that the current density in superconducting nanostrips of $wd\ll \lambda^2$ is identical to that in normal conducting nanostrips of $wd\ll \delta^2/2$, provided that the nanostrip geometry is identical.

\section{Kinetic inductance for nanostrips with turns
\label{app:Lk-turns}} 
Here, we demonstrate analytical formulas of $L_\textrm{k}$ for several superconducting nanostrip geometries. 
These formulas are simply obtained by using Eq.~\eqref{eq:Lk/Rn} and converting from the formulas of the normal resistance $R_\textrm{n}$ presented in Ref.~\onlinecite{Hall_68}. 

\begin{figure}[b]
\includegraphics[width=80mm]{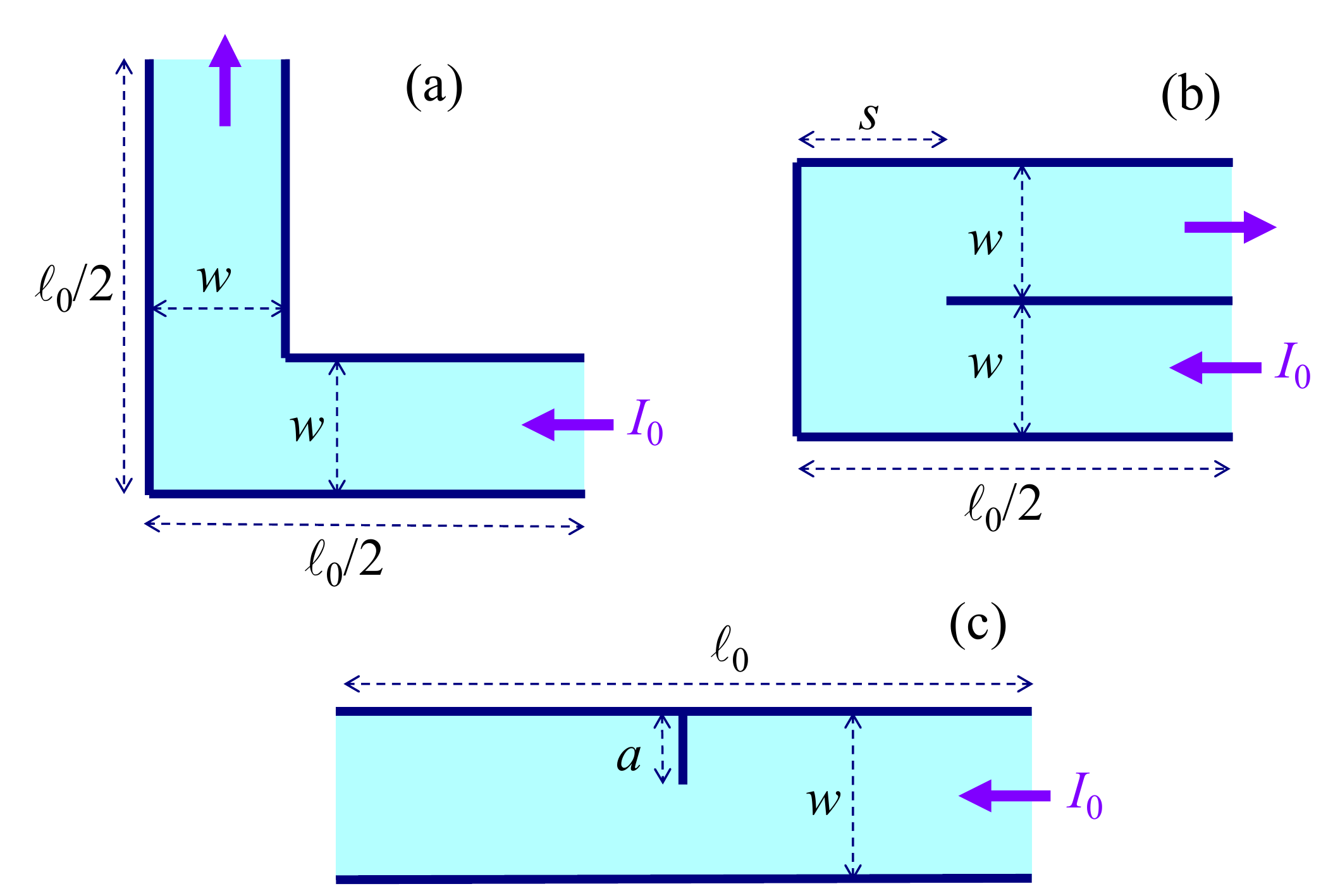}
\caption{%
Superconducting nanostrips of length $\ell_0$ and width $w$ with (a) a $90^{\circ}$ turn, (b) a $180^{\circ}$ turn of turn width $s$, and (c) a thin vertical notch of length $a$.
}
\label{fig:strips_turn-notch}
\end{figure}

For a superconducting nanostrip with a $90^{\circ}$ turn [Fig.~\ref{fig:strips_turn-notch}(a)], kinetic inductance is given by Eq.~\eqref{eq:Lk_general} with 
\begin{equation}
	\alpha=-1-(2/\pi)\ln 2 = -1.44.
\label{eq:Lk-90turn}
\end{equation}

For a superconducting nanostrip with a $180^{\circ}$ turn [Fig.~\ref{fig:strips_turn-notch}(b)], kinetic inductance is given by Eq.~\eqref{eq:Lk_general} as 
\begin{equation}
	\alpha= -\frac{4}{\pi} \ln\left[\sinh\left(\frac{\pi s}{2w}\right)\right], 
\label{eq:Lk-180turn}
\end{equation}
which is identical to Eq.~\eqref{eq:Lk_s-small-medium}.

For a superconducting nanostrip with a thin vertical notch of length $a$ [Fig.~\ref{fig:strips_turn-notch}(c)], kinetic inductance is given by Eq.~\eqref{eq:Lk_general} as 
\begin{equation}
	\alpha= -\frac{4}{\pi} 
		\ln\left[\cos\left(\frac{\pi a}{2w}\right)\right]. 
\label{eq:Lk-notch}
\end{equation}

\section{Transport current, kinetic inductance, and stream function
\label{app:I0-Lk-phi-phi}} 
Because the normal component of the current density vanishes at the sides of a nanostrip, we have ${\bm J}\cdot\hat{\bm n} =\nabla\times(\psi\hat{\bm z})\cdot\hat{\bm n} =(\hat{\bm z}\times\hat{\bm n})\cdot\nabla\psi =0$, which means that the tangential differentiation of $\psi$ at the sides is zero. 
Thus, we obtain the boundary conditions of constant $\psi$ at the sides, $\psi=\psi_1$ at one side and $\psi=\psi_2$ at the other side of a nanostrip, where $\psi_1$ and $\psi_2$ are constants. 

If the current density is normal to the ends of a nanostrip, where a transport current flows in or out, then we have ${\bm J}\cdot(\hat{\bm n}\times\hat{\bm z}) =-\nabla\varphi\cdot(\hat{\bm n}\times\hat{\bm z})=0$, which means that the tangential differentiation of $\varphi$ at the ends is zero. 
Consequently, we obtain the boundary conditions of constant $\varphi$ at the ends, of $\varphi=\varphi_1$ at one end and $\varphi=\varphi_2$ at the other end of a nanostrip, where $\varphi_1$ and $\varphi_2$ are constants.

The net sheet current, $I_0/d$, is calculated as the line integral of the normal component of the current density at the ends, 
\begin{equation}
	\frac{I_0}{d} =\int{\bm J}\cdot\hat{\bm n}\,ds 
	=\int\nabla\psi \cdot(\hat{\bm z}\times\hat{\bm n}) ds 
	=\int\nabla\psi \cdot d{\bm s},  
\label{eq:I0-psi_app}
\end{equation}
and we obtain Eq.~\eqref{eq:I0-psi}. 

Green's theorem in the $xy$-plane,~\cite{Spiegel_81} 
\begin{equation}
	\iint \left( \frac{\partial Q}{\partial x} 
		-\frac{\partial P}{\partial y} \right) dxdy 
	= \oint (Pdx+Qdy), 
\label{eq:Green-PQ}
\end{equation}
with $P=\psi\partial\varphi/\partial x$ and $Q=\psi\partial\varphi/\partial y$ leads to 
\begin{equation}
	\iint \left( \frac{\partial\psi}{\partial x} \frac{\partial\varphi}{\partial y} 
	-\frac{\partial\psi}{\partial y} \frac{\partial\varphi}{\partial x} \right) dxdy 
	= \oint \psi\left( \frac{\partial\varphi}{\partial x} dx 
		+\frac{\partial\varphi}{\partial y} dy  \right). 
\label{eq:Green-psi-phi}
\end{equation}
Substitution of Eqs.~\eqref{eq:Jx} and \eqref{eq:Jy} into the left-hand side of Eq.~\eqref{eq:Green-psi-phi} yields 
\begin{equation}
	\iint\left( J_x^2 +J_y^2 \right) dxdy 
	= \oint \psi d\varphi. 
\label{eq:J^2_int}
\end{equation}
Because $\psi$ is constant at the sides and $\varphi$ is constant at the ends of nanostrips, the right-hand side of Eq.~\eqref{eq:J^2_int} reduces to 
\begin{equation}
	\iint\left( J_x^2 +J_y^2 \right) dxdy 
	= (\psi_2 -\psi_1)\int d\varphi 
	= (\psi_2 -\psi_1) (\varphi_2 -\varphi_1),  
\label{eq:J^2}
\end{equation}
where $\varphi_1$ and $\varphi_2$ are the values of $\varphi$ at the ends of nanostrips. 
Equation~\eqref{eq:J^2} can also be readily derived from 
\begin{equation}
	\iint\left| \frac{d\mathcal{G}}{d\zeta} \right|^2 dxdy 
	= \iint d\varphi d\psi,   
\end{equation}
where $| d\mathcal{G}/d\zeta |^2 =J_x^2+J_y^2$ corresponds to the Jacobian~\cite{Spiegel_81} of the variable transform from $(x,y)$ to $(\psi,\varphi)$. 
Substitution of Eqs.~\eqref{eq:I0-psi} and \eqref{eq:J^2} into Eq.~\eqref{eq:Lk} yields Eq.~\eqref{eq:Lk_phi}.

\section{Simplified expressions of the kinetic inductance of meander nanostrips
\label{app:Lk-simple}} 
Here, we derive simplified expressions for $L_\textrm{k}$ given by Eqs.~\eqref{eq:Lk_s-narrow}, \eqref{eq:Lk_s-medium}, and \eqref{eq:Lk_s-wide} from Eq.~\eqref{eq:Lk_result} with Eqs.~\eqref{eq:k1}, \eqref{eq:c}, \eqref{eq:gamma}, and \eqref{eq:k2}.

Equation~\eqref{eq:Lk_result} for $k_2\ll 1$ or $k_2'=\sqrt{1-k_2^2} \ll 1$ reduces to 
\begin{align}
	L_\textrm{k} &\simeq 
	\frac{\mu_0\lambda^2}{d} \frac{2}{\pi} \ln\left(\frac{4}{k_2}\right) 
	&\text{for } k_2\ll k_2'\simeq 1, 
\label{eq:Lk_k2-small}\\
	L_\textrm{k} &\simeq 
	\frac{\mu_0\lambda^2}{d} \left[\frac{2}{\pi} \ln\left(\frac{4}{k_2'}\right)\right]^{-1}
	&\text{for } k_2' \ll k_2\simeq 1.
\label{eq:Lk_k2p-small}
\end{align}

We are interested in the case where the width of nanostrips is much smaller than the length, $w/\ell_0\ll 1$, and so we have 
\begin{align}
	k_1' &= \sqrt{1-k_1^2}\simeq 4e^{-\pi\ell_0/2w} \ll 1, 
\\
	1-\gamma^2 &\simeq 16e^{-\pi\ell_0/w} \sinh^2 \left(\frac{\pi s}{2w}\right), 
\\
	\frac{k_2}{k_2'} &= \frac{k_1}{k_1'}(1-\gamma^2) 
	\simeq 4e^{-\pi\ell_0/2w} \sinh^2 \left(\frac{\pi s}{2w}\right), 
\label{eq:k2-k2p}
\end{align}
from Eqs.~\eqref{eq:k1}, \eqref{eq:gamma}, and \eqref{eq:k2}. 
Equation~\eqref{eq:k2-k2p} further reduces to 
\begin{align}
	k_2 &\simeq e^{-\pi\ell_0/2w} (\pi s/w)^2 \ll k_2'\simeq 1
	&\text{for } s\ll w,
\label{eq:k2_s-narrow}\\
	k_2 &\simeq e^{-\pi(\ell_0/2-s)/w} \ll k_2'\simeq 1
	&\text{for } w<s<\ell_0/2,
\label{eq:k2_s-medium}\\
	k_2' &\simeq e^{-\pi(s-\ell_0/2)/w} \ll k_2\simeq 1
	&\text{for } s>\ell_0/2. 
\label{eq:k2p_s-wide}
\end{align}
Substitution of Eq.~\eqref{eq:k2_s-narrow} into Eq.~\eqref{eq:Lk_k2-small}, of Eq.~\eqref{eq:k2_s-medium} into Eq.~\eqref{eq:Lk_k2-small}, and of Eq.~\eqref{eq:k2p_s-wide} into Eq.~\eqref{eq:Lk_k2p-small} yields Eqs.~\eqref{eq:Lk_s-narrow}, \eqref{eq:Lk_s-medium}, and \eqref{eq:Lk_s-wide}, respectively. 
Note that Eqs.~\eqref{eq:Lk_s-narrow} and \eqref{eq:Lk_s-medium} are expressed as a single form, 
\begin{equation}
	L_\textrm{k} \simeq \frac{\mu_0\lambda^2}{d} 
		\left\{\frac{\ell_0}{w} -\frac{4}{\pi}\ln
		\left[\sinh\left(\frac{\pi s}{2w}\right)\right]\right\} 
\quad\text{for } s<\ell_0/2 , 
\label{eq:Lk_s-small-medium}
\end{equation}
which is identical to Eq.~\eqref{eq:Lk-180turn}.

\begin{acknowledgments}
I thank John R. Clem for stimulating discussions. 
This work was supported by JSPS KAKENHI Grant Number JP20K05314. 
\end{acknowledgments}

%

\end{document}